\documentclass[aps, 12pt]{revtex4}
\usepackage{latexsym}
\usepackage{graphicx}
\usepackage{youngtab}

\usepackage{amssymb}

\newtheorem{definition}{Definition}
\newtheorem{lemma}{Lemma}
\newtheorem{theorem}{Theorem}

\begin{document}

\title{Exact minimum and maximum of yield with a finite number of decoy light intensities}
\author{Toyohiro Tsurumaru}
\affiliation{Mitsubishi Electric Corporation,
Information Technology R\&D Center\\
5-1-1 Ofuna, Kamakura-shi, Kanagawa,
247-8501, Japan}
\author{\footnote{Now with Thales Laser Japan.}Alexandre Soujaeff and Shigeki Takeuchi}
\affiliation{Research Institute for Electronic Science, Hokkaido University,
Kita-12 Nishi-6, Kita-ku, Sapporo 060-0812, Japan}

\begin{abstract}
In this paper,
for the decoy state method using a finite number of decoy light intensities,
we present an improved upper and lower bounds for the asymptotic yield $y_n$ for $n$-photon states.
In particular if all the light intensities are less than or equal to one,
they are not only a lower or upper bound, but in fact are the exact minimum or maximum.
\end{abstract}

\maketitle

\section{Introduction}
The decoy state method is a technique used in quantum key distribution (QKD)
for determining the possible range of the yield $y_1$ and the error rate $e_1$,
by a statistical test using several different light intensities \cite{Hwang,Wang,LMC}.
Here the yield $y_n$ is the probability that an $n$-photon state emitted by Alice is detected in Bob's apparatus,
and $e_n$ the error rate caused solely by the $n$-photon states.
In this method, Alice first chooses the average photon number of each of her coherent light pulses randomly
out of $\mu_1,\dots,\mu_M$, and Bob records every detection events.
After quantum communications are completed,
Alice reveals the average photon number of each pulse over an authenticated public channel.
Then referring to these data,
Bob calculates the detection rates corresponding to each $\mu_i$,
and estimates a lower bound or the minimum of $y_1$ that is consistent with them.
Similarly, he can also estimate an upper bound or the maximum of $e_1$.

These values are then used to calculate the key generation rate $R$
by plugging them into well-known formulae, e.g.,
$R=Q(\mu) f(E(\mu))H_2(E(\mu))+Q_0(\mu)+Q_1(\mu)[1-H_2(e_1)]$
for the BB84 protocol \cite{Koashi}.
Here $Q(\mu)$ is the overall detection rate in Bob's detector,
and $Q_0(\mu)$, $Q_1(\mu)$ are the contributions to it from the pulses containing zero and one photon respectively.
$E(\mu)$ is the overall error rate, and $H_2(e)$ the binary entropy function
$H_2(e)=-e\log_2 e-(1-e)\log_2(1-e)$,
hence $H_2(E(\mu))$ corresponds to the length of a syndrome consumed to correct bit errors.
The factor $f(E(\mu))$ is inserted to take into account the information rate of practical error correcting codes
which is usually below the Shannon limit.

Lower bounds on $y_1$ with a finite number of decoy intensities have been discussed in many papers
(see, e.g., \cite{Wang,MZL,Wang2,Hayashi} and references therein),
and the best and the most general one is due to Hayashi \cite{Hayashi}.
Adding to these results,
in this paper we present an improved upper and lower bounds $X_n$, $Z_n$ for the asymptotic yield $y_n$.
In particular if all the light intensities $\mu_i$ are less than or equal to one,
$X_n$, $Z_n$ are not only a lower or an upper bound,
but in fact are the exact minimum or the maximum.

The main difference between preceding approaches and ours is as follows.
The original decoy problem is an optimization problem involving an infinite number of variables $y_n$.
In order to reduce the number of variables to finite,
Wang devised a decomposition of a phase-randomized coherent state \cite{Wang},
which was later generalized to the case of an arbitrary number of decoy light intensities by Hayashi \cite{Hayashi};
What they did was to decompose the state $\rho$ sent by Alice as a sum
$\rho=\sum_{n=1}^{N}a_n\rho_n$  of mixed states $\rho_n$.
Then by regarding $a_n$ as independent variables and using a linear-programming-like approach \cite{Kreyszig},
Hayashi presented a general algorithm for obtaining the minimum of $y_1$, which is linear in $a_n$.
At first this method may seem general enough and capable of giving the exact minimum of $y_n$.
So what needs to be improved further?
The answer is that $a_n$ cannot be considered as independent in reality since $\rho_n$ are not
completely distinguishable to each other.
Thus by regarding that way they give Eve more power than she actually has,
and there is no guarantee that the obtained minimum is also that of the original problem
involving an infinite number of $y_n$.

In contrast, in this paper we present a method for finding the minimum of $y_1$ without reducing the variable concerned,
i.e., we treat all $y_n$'s independently as in the original decoy method problem.
The key observation here is that when regarding variables $y_n$ as an infinite-dimensional vector $y$,
the difference Eve can make to $y$ without being noticed by Alice or Bob
can be expanded by a set of basis vectors $w^{(m)}$,
each of which is written in a simple form with the Schur polynomials.

What is remarkable about our result is that the configuration of $y_n$ leading to the smallest $y_1$ varies
depending on whether the number $M$ of decoy light intensities (including the signal) is even or odd.
The analysis is especially simple for $M$ even and $\mu_i\le1$;
Because of the positivity of the Schur polynomials,
it is readily seen that $y_1$ is minimized when $y_n=0$ for $n>M$
and that the problem is automatically reduced to that involving only a finite number of variables;
$y_1,\dots,y_M$.
Thus by simply inverting a matrix, the minimum of $y_1$ is expressed in an explicit and simple form.
On the other hand for $M$ odd, 
the analysis turns out to be somewhat more complicated,
however, we can still specify the configuration that corresponds to the smallest
$y_1$ and write down an explicit algorithm for finding it out within a finite number of steps.

The paper is organized as follows.
In Section \ref{sec:main_result}
we define our problem of the decoy state method and present our main result.
Section \ref{sec:M_even} analyzes configuration $X_n$ which is in particular
useful in determining the minimum of $y_1$ when an even number $M$ of decoy light intensities are used.
Subsequently in Section \ref{sec:M_odd} we discuss the properties of $Z_n$ which is useful for $M$ odd.
Finally we conclude in Section \ref{sec:conclusion}.


\section{Setups and Statement of the main result}\label{sec:main_result}
\subsection{Decoy method}
Throughout the paper, for the sake of simplicity,
we consider the case where $y_0$ is already known precisely by using vacuum decoy states,
and we discuss the minimum and maximum of $y_1$ under the condition that
\begin{equation}
Q_+(\mu_i):=e^{-\mu_i}\sum_{n=1}^\infty \frac{\mu_i^n}{n!}y_n
\label{eq:constraint_Q_positive}
\end{equation}
is satisfied for $i=1,\dots,M$.
Note, however, our analysis in the subsequent sections is equally valid even without vacuum decoy states.
$Q_+(\mu_i)$ appearing in (\ref{eq:constraint_Q_positive}) denotes the contribution
from non-zero photon number state to the detection rate in Bob's detector,
i.e., $Q_+(\mu):=Q(\mu)-e^{-\mu}y_0$.
Being a probability, each $y_n$ is of course constrained as
\begin{equation}
0\le y_n\le1\ \ {\rm for\ all}\ \ n\ge1.
\label{eq:complete_boundary_condition}
\end{equation}
The explicit form of the detection rate $Q_+(\mu)$ depends on the physical model
that one employs for describing the quantum channel.
In this paper, we assume that in the absence of Eve,
the yield takes the value $y_n=q_n$ with
\begin{eqnarray}
q_n&:=&A\eta_n+B,\label{eq:error_model}\\
\eta_n&:=&1-(1-\eta)^n,\nonumber
\end{eqnarray}
and that each parameters are conditioned as
\begin{equation}
0\le A\le1,\ \ 0\le B\le\eta\le1/10.
\label{eq:parameter_domain}
\end{equation}
Here $\eta$ is the channel transmission rate including the quantum efficiency of Bob's detector,
and $B$ is roughly the dark count rate $p_{\rm dark}$.
Note that for practical QKD systems, (\ref{eq:parameter_domain}) is not really a restriction;
$\eta$ is already around 0.1 at 0km due to the detector efficiency.
On the other hand for sufficiently small $\eta$,
we have $B\le \frac12Q(\mu_i)E(\mu_i)\simeq \frac12\eta\mu_i E(\mu_i)\le\frac12\eta\mu_i$.
Thus with the signal light intensity (say $\mu_1$) normally being around 0.5 or less,
$B\le \eta$ is automatically satisfied.

According to Lo et al. \cite{LMC} and Hayashi \cite{Hayashi},
these are $A=1$, $B=p_{\rm dark}$,
from which we have
\[
Q(\mu_i)=1-\exp(-\eta\mu_i)+p_{\rm dark},
\]
whereas in some other references (e.g., \cite{Koashi}),
slightly different models such as $A=1-p_{\rm dark}$ and $B=p_{\rm dark}$ are used
\footnote{In fact the former model can only be considered as an approximation valid for smaller values of $n$,
since $y_n$ exceeds one for large enough $n$.
Hence in a strict sense,
one needs to introduce a cut off $N$ such that $y_n$ may be neglected for $n\ge N$,
or use different definitions such as the latter.}.

The decoy state method is similarly effective in lower bounding the error rate $e_1$
from pulses containing a single photon;
By recording the overall error rate $E(\mu_i)$ for each decoy light intensity $\mu_i$
and using the relation
\begin{equation}
Q(\mu_i)E(\mu_i)-\frac12e^{-\mu_i}y_0=e^{-\mu_i}\sum_{n=1}^\infty \frac{\mu_i^n}{n!}b_n
\label{eq:constraint_e}
\end{equation}
with $b_n:=y_ne_n$,
one can determine the range of $b_1=y_1e_1$.
This case can also be treated with (\ref{eq:error_model}) by redefining parameters $A$, $B$.
For instance in \cite{LMC, Hayashi}, the value on the left hand side of Eq. (\ref{eq:constraint_e}) takes
the form
\[
Q(\mu_i)E(\mu_i)=e_{\rm det}\left(1-\exp(-\eta\mu_i)\right)+\frac12p_{\rm dark},
\]
which corresponds to $A=e_{\rm det}$, $B=p_{\rm dark}/2$.
A slightly different error models are also used, e.g., in \cite{Koashi}.
In what follows we do not distinguish between all these cases,
whether of yields or of error rates,
and analyze them on an equal footing as an optimization problem regarding Eq. (\ref{eq:error_model})
with given values of $A$, $B$ satisfying (\ref{eq:parameter_domain}).

\subsection{Main result}\label{sec:Main_result}
Under these settings,
we present upper and lower bounds on $y_n$ in terms of quantities $X_n$ and $Z_n$;
for any $M$ and $n<M$,
\begin{eqnarray}
X_n\le y_n\le Z_n&\ &{\rm if}\ M-n\ {\rm is\ odd},\label{eq:statement_even}\\
Z_n\le y_n\le X_n&\ &{\rm if}\ M-n\ {\rm is\ even},\label{eq:statement_odd}
\end{eqnarray}
where $X_n$ are expressed in a simple form (see Eq. (\ref{eq:y_bar_expicit_form})).
For instance, $X_1$ takes the form
\begin{equation}
X_1=\sum_{i=1}^M\frac{\exp(\mu_i)Q_+(\mu_i)}{\mu_i}\prod_{j=1,j\ne i}^M\frac{\mu_j}{\mu_j-\mu_i}.
\label{eq:y_1_explicit_form}
\end{equation}
On the other hand $Z_n$ cannot be written in a simple form as $X_n$,
however,
as we shall demonstrate in Section \ref{sec:M_odd},
they can always be obtained by a numerical calculation within a finite number of steps.

In addition, it can be shown that at least when $\mu_i\le1$,
Eve can actually attain $y_n=X_n$ and $y_n=Z_n$ in (\ref{eq:statement_even}) and (\ref{eq:statement_odd}).
Hence they are not only a lower (resp. upper) bound, but in fact are the minimum (resp. maximum) of $y_1$.

In order to demonstrate how effective our approach is,
take a typical set of experimental parameters, e.g., $A=1$, $\eta=10^{-2}$, $B=p_{\rm dark}=10^{-5}$,
$M=3$, and $(\mu_1,\mu_2,\mu_3)=(0.07, 0.2, 0.5)$.
It turns out that $Z_1=0.993\times10^{-2}\le y_1\le1.003\times10^{-2}=X_1$,
where the yield in the absense of Eve is $y_1=q_1=1.001\times10^{-2}$.
Hence by using only four decoy light intensities including vacuum,
we can determine $y_1$ within accuracy of less than one percent.


\section{Minimum of $y_1$ for $M$ even}\label{sec:M_even}
First in this section,
we discuss the property of $X_n$ as lower or upper bounds as stated in Sec. \ref{sec:Main_result}.
This is in particular useful in determining the minimum of $y_1$
when there are an even number of constraints, that is, for $M$ even.

For $M=2$ Hwang \cite{Hwang} pointed out that Eve's best attack strategy is to set $y_n=0$ for all $n\ge3$,
and hence the problem is reduced to solving an linear equation of $y_1$, $y_2$.
Here we shall show that this can in fact be generalized to any even value of $M$,
i.e., in order to obtain the minimum $y_1$, it suffices to set $y_n=0$ for all $n>M$
and calculate $y_1$ compatible with
\begin{equation}
\left(
\begin{array}{cccc}
\mu_1&\mu_1^2&\cdots&\mu_1^M\\
\mu_2&\mu_2^2&\cdots&\mu_2^M\\
\vdots&\vdots&\ddots&\vdots\\
\mu_M&\mu_M^2&\cdots&\mu_M^M\\
\end{array}
\right)
\left(
\begin{array}{c}
y_1/1!\\
y_2/2!\\
\vdots\\
y_M/M!\\
\end{array}
\right)
=
\left(
\begin{array}{c}
\exp(\mu_1)Q_+(\mu_1)\\
\exp(\mu_2)Q_+(\mu_2)\\
\vdots\\
\exp(\mu_M)Q_+(\mu_M)
\end{array}
\right)
\label{eq:M_constraints}
\end{equation}
by inverting the Vandermonde matrix.
For the rest of the paper, we denote the solution $y_n$ to Eq. (\ref{eq:M_constraints}) as $X_n$.
For $n>M$, we set $X_n=0$ formally for later convenience.

\begin{theorem}
\label{th:M_even}\ 

\begin{itemize}
\item For $M$ even,
$X_1$ is a lower bound of $y_1$ which is consistent with Eq. (\ref{eq:constraint_Q_positive}).
\item More generally, for any $M$ and any $n\le M$,
$X_n$ is a lower (resp. upper) bound of $y_n$ if $M-n$ is an odd (resp. even) number.
\item
If $\mu_1,\dots,\mu_M\le 1$, we have $0\le X_n\le1$ for all $n$.
That is, Eve can actually achieve $y_n=X_n$.
Hence $X_n$ is not only a lower (resp. upper) bound,
but is also the minimum (resp. maximum) of $y_n$ for $M-n$ odd (resp. even).
\end{itemize}
\end{theorem}
The proof will be given in Section \ref{sec:proof_th_1}.
Using Cramer's rule, the solution $X_n$ to Eq. (\ref{eq:M_constraints}) can be expressed explicitly as
\begin{equation}
\frac{X_n}{n!}=\left|
\begin{array}{ccccccc}
\mu_1&\cdots&\mu_1^{n-1}&\exp(\mu_1)Q_+(\mu_1)&\mu_1^{n+1}&\cdots&\mu_1^M\\
\vdots&\vdots&\vdots&\vdots&\vdots&\vdots&\vdots\\
\mu_M&\cdots&\mu_M^{n-1}&\exp(\mu_M)Q_+(\mu_M)&\mu_M^{n+1}&\cdots&\mu_M^M
\end{array}
\right|
/D(\mu_1,\dots,\mu_M),
\label{eq:y_bar_expicit_form}
\end{equation}
\begin{equation}
D(\mu_1,\dots,\mu_M):=\left(\prod_{i=1}^M \mu_i\right)\Delta(\mu_1,\dots,\mu_M)
\label{eq:def_D}
\end{equation}
with $\Delta(\mu_1,\dots,\mu_M)$ being the Vandermonde determinant
\begin{equation}
\Delta(\mu_1,\dots,\mu_M)
:=
\left|
\begin{array}{ccc}
1&\cdots&\mu_1^{M-1}\\
\vdots&\ddots&\vdots\\
1&\cdots&\mu_M^{M-1}
\end{array}
\right|.
\label{eq:def_Vandermonde}
\end{equation}
In particular, $X_1$ takes the form of Eq. (\ref{eq:y_1_explicit_form}).

\subsection{Mathematical preliminary}
As a preliminary to the proof of Theorem \ref{th:M_even},
we define the Schur polynomials $s_\lambda$ (see, e.g., Ref. \cite{Fulton1,Fulton2}) and difference vectors $w^{(m)}_n$.
\begin{definition}\label{def:Schur_polynomial}
Choose an integer partition $\lambda=(\lambda_1,\dots,\lambda_k)$
satisfying $\lambda_1\ge\lambda_2\ge\cdots\ge\lambda_k>0$, and $k\le M$.
For $n>k$, set $\lambda_n=0$ formally.
The Schur polynomial $s_\lambda$ in variables $\mu_1,\dots,\mu_M$
is defined as
\[
s_\lambda(\mu_1,\dots,\mu_M):=\left|
\begin{array}{cccc}
\mu_1^{\lambda_M}&\mu_1^{1+\lambda_{M-1}}&\cdots&\mu_1^{M-1+\lambda_1}\\
\mu_2^{\lambda_M}&\mu_2^{1+\lambda_{M-1}}&\cdots&\mu_2^{M-1+\lambda_1}\\
\vdots&\vdots&\vdots&\vdots\\
\mu_M^{\lambda_M}&\mu_M^{1+\lambda_{M-1}}&\cdots&\mu_M^{M-1+\lambda_1}
\end{array}
\right|
\ /\ \Delta(\mu_1,\dots,\mu_M),
\]
where $\Delta(\mu_1,\dots,\mu_M)$ is the Vandermonde determinant defined in Eq. (\ref{eq:def_Vandermonde}).
\end{definition}
For example, if the partition $\lambda$ is empty, i.e., $\lambda_1=\lambda_2=\cdots=0$,
both the numerator and the denominator equal $\Delta(\mu_1,\dots,\mu_M)$ and we have $s_{\emptyset}=1$.
For $\lambda=(1,1,\dots,1)$ with $1$ repeating $M$ times
$s_{(1,1,\cdots,1)}=\prod_{i=1}^M \mu_i$.
In what follows, we denote integer partitions with greek letters $\lambda,\alpha,\dots$
with the only exception of $\mu$ that is used for average photon numbers.

Now using $s_\lambda$ thus defined, we consider difference vectors $\Delta y=(\Delta y_1,\Delta y_2,\dots)$
to $y=(y_1,y_2,\dots)$ which preserve the constraint (\ref{eq:constraint_Q_positive}).
In other words $\Delta y$ are those vectors satisfying
\begin{equation}
\sum_{n=1}^\infty \frac{\mu_i^n}{n!}\Delta y_n=0
\label{eq:def_Delta}
\end{equation}
for all $1\le i\le M$.
Hence if $y$ is a solution to (\ref{eq:constraint_Q_positive}),
$y+\Delta y=(y_1+\Delta y_1, y_2+\Delta y_2, \dots)$ is also a solution when disregarding the constraints $0\le y_n\le 1$.
The set of vectors $W:=\{\Delta y\ {\rm satisfying\ Eq.\ (\ref{eq:def_Delta})}\}$
clearly forms a subspace of the vector space $V$ consisting of all vectors \footnote{%
To be precise,
we assume that $V$ consists of $v=(v_1,v_2,\dots)$ satisfying $\sum_{n=1}^\infty \mu^nv_n/n!<\infty$.
}.
For our present purposes, it is convenient to choose the following non-orthogonal basis for $W$.
\begin{definition}\label{def:Delta_y}
We define a set of vectors $w^{(m)}=(w^{(m)}_1,w^{(m)}_2,\dots)$ labeled by $m>M$ as
\begin{equation}
w^{(m)}_n=
\left\{
\begin{array}{cl}
\displaystyle{(-1)^{M-n+1}\frac{n!}{m!}s_{\alpha(m-M,M-n)}(\mu_1,\dots,\mu_M)}& for\ \ n<M,\\
1& for\ \ n=m,\\
0&otherwise,
\end{array}
\right.
\label{eq:def_w}
\end{equation}
where $\alpha$ denotes an integer partition
$\alpha(a,b):=(a,1,1,\dots,1)$ with $1$'s repeating $b$ times.
\end{definition}

\begin{lemma}\label{lm:w_completeness}
Vectors $w^{(m)}$ form a linear basis of $W$.
That is, $w^{(m)}$ are solutions to Eq. (\ref{eq:def_Delta}),
and conversely,
any solution to Eq. (\ref{eq:def_Delta}) can be uniquely expressed as
a superposition of $w^{(m)}$ as
\begin{equation}
\Delta y=\sum_{m=M+1}^\infty \Delta y_mw^{(m)}.
\end{equation}
\end{lemma}

The proof is given in Appendix \ref{app:proof_lemma_1}.
With the help of this lemma,
we see that given any solution $y=(y_1,y_2,\dots)$ to Eq. (\ref{eq:constraint_Q_positive}),
$X-y$ is written uniquely as a superposition of $w^{(m)}$ as
\begin{equation}
y_n-X_n=\sum_{m=M+1}^\infty w_n^{(m)}(y_m-X_m)=\sum_{m=M+1}^\infty w^{(m)}_ny_m.
\label{eq:difference_expanded}
\end{equation}
We will use this relation repeatedly in the following sections.

\subsection{Proof of Theorem 1}\label{sec:proof_th_1}
\begin{figure}[htbp]
\begin{center}
\includegraphics[width=100mm]{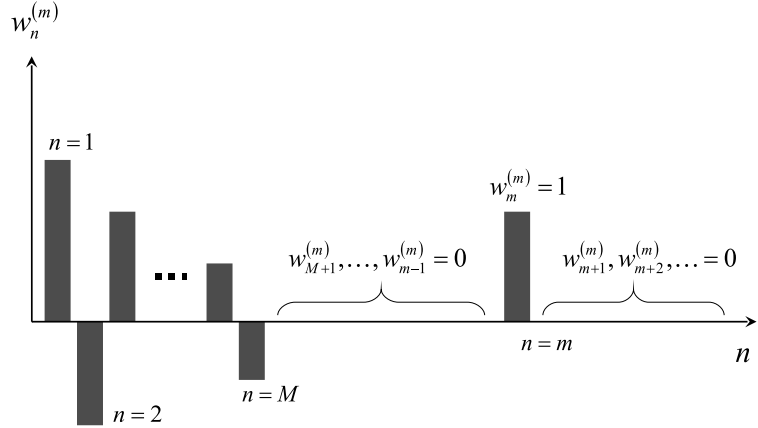}
\end{center}
\caption{Image of $w^{(m)}_n$ for $M$ even.
The element changes signs as $n$ increases from $1$ to $M$ and then to $m$.}
\label{fig:one}
\end{figure}

In this subsection we will prove Theorem 1,
but before going into details, let us give an intuitive explanation.
Eve's goal is to minimize $y_1$ while keeping the measured value of $Q_+(\mu_i)$ intact
so that her attack will not be noticed by Alice and Bob.
Hence the difference $\Delta y$ she makes to the yield $y$ must satisfy (\ref{eq:def_Delta}),
and as we have seen in Lemma \ref{lm:w_completeness},
it can always be considered as a sum of the basis vector $w^{(m)}$.
Now note that the Schur polynomial $s_{\alpha(a,b)}$ being always positive in Eq. (\ref{eq:def_w}),
the element of $w^{(m)}_n$ alternates its signs with as $n$ increases as $n=1,\dots,M$ and $m$.
In pariticular if $M$ is even, both $w^{(m)}_1$ and $w^{(m)}_m$ are positive for any $m$
(see Fig. \ref{fig:one}).
Thus we see that minimizing $y_n$ for $n>M$, or equivalently, taking $\Delta y_n\le0$ will always decrease $y_1$.
As a result, the best configuration for Eve turns out to be the one with $y_n=0$ for all $n>M$,
i.e., $X_n$.

\begin{lemma}\label{le:minimum}
For $M$ even,
$X_1$ is a lower bound on $y_1$ of Eq. (\ref{eq:constraint_Q_positive}).
More generally, for any $n\le M$,
if $M-n$ is odd (resp. even),
$X_n$ is a lower (resp. upper) bound on $y_n$.
\end{lemma}

{\it Proof:}
Since the proof is essentially the same for all cases,
we consider here only the case of $n=1$ and $M$ being even.
During the proof,
we suppress the constraint $0\le y_n\le1$ for $n=1,\dots,M$ and let them take an arbitrary value.
For $n>M$ we still require $y_n\ge0$.
Then in fact $X_1$ is the minimum of $y_1$ under these requirements,
and is also a lower bound under the full constraint (\ref{eq:complete_boundary_condition}).
This can be seen by looking at the $n=1$ element of Eq. (\ref{eq:difference_expanded});
Given an arbitrary solution $y=(y_1,y_2,\dots)$ to Eq. (\ref{eq:constraint_Q_positive}),
the difference of its first element with $X$'s takes the form
\[y_1-X_1=\sum_{m=M}^\infty w^{(m)}_1y_m.\]
Obviously this is always greater than or equal to zero because $y_m\ge0$ and $w^{(m)}_1>0$ for $M$ even.
Hence $X_1$ is the minimum of $y_1$ under the requirements that we introduced at the beginning.
This completes the proof.

From this proof
we see that if all of $X_1,\dots,X_M$ satisfy $0\le X_n\le 1$
for a particular choice of $A$ and $B$,
they are indeed the true minima (resp. maxima) under the full constraints
(\ref{eq:complete_boundary_condition}).
One can always verify this by numerical calculations, and doing so may be useful in practice.
However, we can in fact verify it analytically for a sufficiently wide range of parameters.

\begin{lemma}\label{lm:boundary}
For $\mu_1,\dots,\mu_M\le1$, we have $0\le X_n\le 1$ for all $n$.
\end{lemma}

{\it Proof:}
Substituting $y_n=q_n$ in Eq. (\ref{eq:difference_expanded}) and using Eq. (\ref{eq:def_w}),
we obtain
\begin{eqnarray}
\frac{X_n}{n!}&=&\frac{q_n}{n!}+(-1)^{M-n}I_n,\label{eq:y_n_extended}\\
I_n:&=&\sum_{m=M+1}^\infty \frac{q_m}{m!}s_{\alpha(m-M,M-n)}(\mu_1,\dots,\mu_M)\nonumber
\end{eqnarray}
for $n\le M$.
According to the positivity of $q_n$ and the Schur polynomials $s_\lambda$,
we have $I_n\ge0$.
From this it is immediate that $y_n\ge0$ for $M-n$ even,
and $y_n\le1$ for $M-n$ odd.
No that so far we did not use the condition $\mu_i\le1$.

On the contrary, in order to see $y_n\le1$ for $M-n$ even and $y_n\ge0$ for $M-n$ odd,
we need to bound $I_n$ from above using $\mu_i\le1$.
By inequality (\ref{eq:s_lambda_bound}) and $\eta_n\le n\eta$,
\begin{eqnarray}
I_n&\le& \sum_{m=M+1}^\infty \frac{Am\eta+B}{m!}\mu_M^{m-n}\frac{(m-n-1)!}{(M-n)!(m-M-1)!}\nonumber\\
&\le&\sum_{m=M+1}^\infty \frac{A\eta+B}{(m-1)!}\mu_M^{m-n}\frac{(m-n-1)!}{(M-n)!(m-M-1)!}\nonumber\\
&=&\frac{\mu_M^{M-n+1}(A\eta+B)}{M!}\sum_{k=0}^\infty 
\frac{A\eta+B}{k!}\,\mu_M^k\,\frac{M\cdots(M-n+1)}{(k+M)\cdots(k+M-n+1)},\label{eq:I_n_upperbound_origin}
\end{eqnarray}
thus for $\mu_i\le1$,
\begin{equation}
I_n\le\frac{A\eta+B}{M!}\sum_{k=0}^\infty \frac1{k!}=\frac{e\left(A\eta+B\right)}{M!}
\label{eq:I_n_upperbound_1}
\end{equation}
for all $M$ and $n\le M$.
On the contrary, inequality (\ref{eq:I_n_upperbound_origin}) for $M=2$ and $n=1$ in particular yields
\begin{eqnarray}
I_1&\le&(A\eta+B)\sum_{k=0}^\infty\frac{k+1}{(k+2)!}\nonumber\\
&=&(A\eta+B)\sum_{k=0}^\infty\left(\frac1{(k+1)!}-\frac1{(k+2)!}\right)\nonumber\\
&=&A\eta+B.
\label{eq:I_n_upperbound_M2n1}
\end{eqnarray}
Therefore, combining (\ref{eq:I_n_upperbound_1}) and (\ref{eq:I_n_upperbound_M2n1}) we obtain for $M-n$ odd,
\begin{equation}
I_n\le\frac{A\eta+B}{(M-1)!}.
\label{eq:I_n_upperbound_m_minus_n_odd}
\end{equation}

Now by using (\ref{eq:I_n_upperbound_1}) for $M-n$ even, or $n=M,M-2,\dots>0$, we have
\[
X_n\le q_n+\frac{n!\,e}{M!}(A\eta+B)\le\left(1+\frac{n!\,e}{M!}\right)(A\eta+B)\le(1+e)(A\eta+B).
\]
The second inequality follows from $\eta_n\le n\eta$ and thus $q_n/n!\le A\eta+B$.
Then using condition (\ref{eq:parameter_domain}) we see
\[X_n\le(1+e)(A\eta+B)\le(1+e)2\eta<1.\]
for all even $n\le M$.
Similarly for $M-n$ odd, or $n=M-1,M-3,\dots>0$,
by using (\ref{eq:I_n_upperbound_m_minus_n_odd}) we find
\[\frac{X_n}{n!}\ge\frac{q_n}{n!}-\frac{A\eta+B}{(M-1)!}
\ge \frac{1}{(M-1)!}\left(q_{M-1}-(A\eta+B)\right).
\]
In the second inequality,
we used the fact that $q_n/n!$ is monotinically decreasing in $n$.
Since
\[
q_{M-1}-(A\eta+B)=A(\eta_{M-1}-\eta)\ge0
\]
for $M\ge2$, we have finally $X_n\ge0$ for $M-n$ odd.
This completes the proof.


\section{Minimum of $y_1$ for $M$ odd}\label{sec:M_odd}
For $M$ odd as well, by using a similar argument as used in the previous section,
the configuration $y_n$ giving the minimum value of $y_1$ can be determined if $\mu_1,\dots,\mu_M\le1$.
In what follows we denote this configuration as $Z=(Z_1,Z_2,\dots)$.
$Z$ includes a set of variables $(L,a)$ that can be specified (as far as we know) 
only by numerical calculations,
and cannot be written in a simple form as Eq. (\ref{eq:y_1_explicit_form}).
Still, as shown below, it can always be determined within a finite number of steps.

\subsection{Definition of $Z$}
\begin{figure}[htbp]
\begin{center}
\includegraphics[width=100mm]{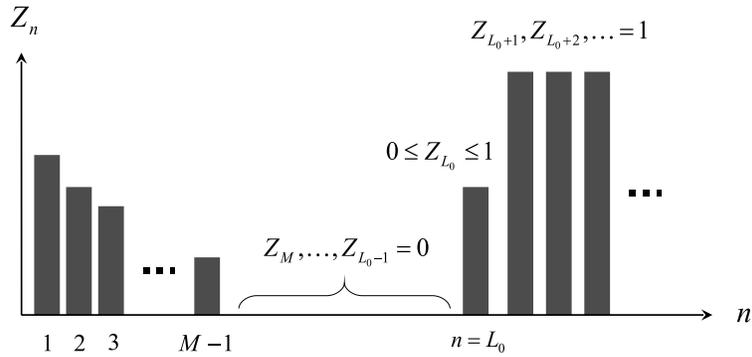}
\end{center}
\caption{Configuration of $Z$ achieving the minimum $y_1=Z_1$ for $M$ odd.
It is a solution to Eq. (\ref{eq:constraint_Q_positive}),
such that (a) $Z_M=0$ 
(b) There exists a value $L_0(>M)$; and $Z_n=0$ for $M\le n<L_0$, $Z_n=1$ for $n>M$,
and $0\le Z_{L_0}=a_0\le1$ are satisfied.
(c) The constraints $0\le y_n\le1$ are suppressed for $Z_1,\dots,Z_{M-1}$ and they can take an arbitrary value.
}
\label{fig:two}
\end{figure}

In this subsection we define what the configuration $Z$ looks like in two steps;
First we give a configuration $z$ involving parameters $L$, $a$ and then define $Z$ as its special case.
\begin{definition}\label{def:define_small_z}
For a given set of an integer $L>M$ and a real number $0<a\le1$,
$z(L,a)=(z_1(L,a),z_2(L,a),\dots)$ is configuration of the yield $y$,
and is a solution to Eq. (\ref{eq:constraint_Q_positive})
satisfying the following conditions (see Fig. \ref{fig:two}).
\begin{itemize}
\item $z_n=0$ for $M<n<L$ and $z_n=1$ for $L<n$.
\item $z_L=a$.
\item 
Constraint (\ref{eq:complete_boundary_condition}) is relaxed for $n=1,\dots,M-1$.
That is, $z_1,\dots,z_{M-1}$ can take an arbitrary value.
\end{itemize}
\end{definition}
Let us supplement this definition.
As we have seen in Eq. (\ref{eq:difference_expanded}),
once $z_{M+1},z_{M+2},\dots$ are all fixed,
$z_1,\dots,z_M$ are uniquely determined as
\begin{equation}
\frac{z_M(L,a)}{M!}=\frac{X_M}{M!}-\frac{a}{L!}s_{(L-M)}(\mu_1,\dots,\mu_M)
-\sum_{m=L+1}^\infty\frac1{m!}s_{(m-M)}(\mu_1,\dots,\mu_M).
\label{eq:z_M_expanded}
\end{equation}
The third item of Definition \ref{def:define_small_z} means that
we do not care whether the value thus obtained satisfy $0\le z_1(L,a),\dots,z_M(L,a)\le 1$ or not.
Using this $z(L,a)$, we now define $Z$.

\begin{definition}\label{def:define_large_Z}
Configuration $Z$ is $z(L,a)$ with the smallest $L$ and the largest $a$
satisfying $z_M(L,a)\ge0$.
In what follows we denote such $(L,a)$ as $(L_0, a_0)$,
and thus $Z=z(L_0,a_0)$.
\end{definition}

In order for this definition to make sense,
we need to guarantee the existence and the uniqueness of $(L_0,a_0)$ for an arbitrary choice of $A$ and $B$.
To see this,
it is convenient to order the pairs $(L,a)$ such that
$(L_1,a_1)>(L_2,a_2)$ if either (i) $L_1>L_2$ or (ii) $L_1=L_2$ and $a_1<a_2$.
In terms of this ordering,
$(L_0,a_0)$ just corresponds to the smallest $(L,a)$ satisfying $z_M(L,a)\ge0$.
By definition, pairs $(L,a)$ are bounded from below by $(M+1,1)$,
and as one can see from (\ref{eq:z_M_expanded}),
$z_M(L,a)$ is monotonically increasing with respect to $(L,a)$.
Hence $(L_0,a_0)$ can obviously be determined uniquely.

We can also show that $L$ is finite.
Indeed if $z_M(L,1)<0$ for any finite $L$,
we would have $X_M=\lim_{L\to\infty} z_M(L,0)\le0$.
However, this would never happen as we have seen in the first paragraph of the proof of Lemma \ref{lm:boundary}.

With this $Z$ the following theorem holds.

\begin{theorem}
\label{th:M_odd}\ 
\begin{itemize}
\item For $M$ odd,
$Z_1$ is a lower bound of $y_1$ which is consistent with Eq. (\ref{eq:constraint_Q_positive}).
\item More generally, for any $M>1$ and $n\le M$,
$Z_n$ is a lower (resp. upper) bound of $y_n$ if $M-n$ is even (resp. odd).
\item
If $\mu_1,\dots,\mu_M\le 1$, we have $0\le Z_n\le1$ for all $n$.
That is, Eve can actually achieve $y_n=Z_n$.
Hence $Z_n$ is not only a lower (resp. upper) bound,
but is also the minimum (resp. maximum) of $y_n$ for $M-n$ even (resp. odd).
\end{itemize}
\end{theorem}
The proof will be given in Section \ref{sec:proof_th_odd}.

\subsection{An algorithm for finding $Z_1$}\label{sec:algorithm_Z_n}
Next in order to demonstrate that $Z_1$ can be actually obtained within finite steps,
we present an algorithm for calculating it.
First note that for given $L$ and $a$,
by plugging $z(L,a)$ in Eq. (\ref{eq:constraint_Q_positive})
we obtain
\[
G(\mu;\,L,a)=\sum_{n=1}^M\frac{\mu^n}{n!}z(L,a)
\]
with
\[
G(\mu;\,L,a):=e^\mu Q_+(\mu)-\left(e^\mu-\sum_{n=0}^L\frac{\mu^n}{n!}+\frac{\mu^L}{L!}a\right).
\]
Then by using Cramer's rule as in Eq. (\ref{eq:y_bar_expicit_form}),
$z_n(a,L)$ for $1\le n\le M$ is given as
\begin{equation}
z_n(L,a)=\frac1{n!}\left|
\begin{array}{ccccccc}
\mu_1&\cdots&\mu_1^{n-1}&G(\mu_1\,;\,L,a)&\mu_1^{n+1}&\cdots&\mu_1^M\\
\vdots&\vdots&\vdots&\vdots&\vdots&\vdots&\vdots\\
\mu_M&\cdots&\mu_M^{n-1}&G(\mu_M\,;\,L,a)&\mu_M^{n+1}&\cdots&\mu_M^M\\
\end{array}
\right|
/D(\mu_1,\dots,\mu_M)
\label{eq:z_n}
\end{equation}
with $D(\mu_1,\dots,\mu_M)$ defined in (\ref{eq:def_D}).
Now that we have got rid of all inifinite series,
$Z$ can be obtained numerically as follows.

\noindent{\bf Algorithm}
\begin{enumerate}
\item Calculate $z_M(M+1,1)$ using Eq. (\ref{eq:z_n}).
If $z_M(M+1,1)\ge0$, let $L_0=M+1$, $a_0=1$ and go to Step \ref{it:obtain_Z_1}.
\item Let $L=M+1$.
\item \label{it:repeat}
If $z_M(L,1)<0$, let $L=L+1$ and go to Step \ref{it:repeat}.
Otherwise let $L_0=L$
and find the root $a_0$ of $z_M(L_0,a_0)=0$.
\item \label{it:obtain_Z_1}
Calculate $Z_1=z_1(L_0,a_0)$ and stop.
\end{enumerate}

Note that we need $z_M(M+1,1)$ and $z_M(L,1)$ in Steps 1 and 3 only in order to check their plus or minus sign.
Hence when actually running the algorithm,
one may omit the division by $D(\mu_1,\dots,\mu_M)$ appearing in Eq. (\ref{eq:z_n})
by ordering $\mu_i$, for example, as $\mu_1<\cdots<\mu_M$.

This algorithm stops within finite steps,
since $L$ is finite as shown in the previous subsection.
Moreover, when $\mu_i\le1$, $L_0$ is bounded from above as $L_0(L_0-M)!\le Me/q_M$,
as shown in Appendix \ref{seq:upper_bound_L_0}.
Hence, e.g. for $M=3$, $\eta=10^{-3}$ and $A=1$, we have $L_0\le10$.

\subsection{Proof of Theorem \ref{th:M_odd}}\label{sec:proof_th_odd}
In this subsection we prove Theorem \ref{th:M_odd}.
As in the previous section,
we first show that $Z_n$ are a lower or upper bound,
and after that we demonstrate that $Z_n$ satisfy constraint (\ref{eq:complete_boundary_condition}) if $\mu_i\le1$.

\begin{lemma}\label{le:lower_bound_Z}
For $M$ odd, $Z_1$ is a lower bound on $y_1$.
More generally for any $M$ and any $n\le M$,
if $M-n$ is even (resp. odd),
$Z_n$ is a lower (resp. upper) bound on $y_n$.
\end{lemma}

{\it Proof:}
Since the proof is essentially the same for all cases,
we here consider only the case of $n=1$ and $M$ odd.
During the proof we suppress constraint (\ref{eq:complete_boundary_condition}) for $n=1,\dots,M-1$
and let $y_1,\dots,y_{M-1}$ take an arbitrary value.
For $m\ge M$ we still assume $0\le y_m\le1$.
Again, by showing that $Z_1$ is the minimum of $y_1$ with these requirements,
we prove that it is a lower bound under the complete set of constraints (\ref{eq:complete_boundary_condition}).
As in the proof of Lemma \ref{le:minimum},
the difference between any solution $y=(y_1,y_2\dots)$ and $Z$
can be expanded as in Eq. (\ref{eq:difference_expanded}).
Thus the constraint $y_M\ge0$ yields
\[
y_M=X_M+\sum_{m=M+1}^\infty w^{(m)}_{M}y_m\ge0,
\]
which can be rewritten by using Eq. (\ref{eq:def_w}) as
\begin{equation}
\frac{X_M}{M!}\ge\sum_{m=M+1}^\infty\frac{y_m}{m!}s_{(m-M)}(\mu_1,\dots,\mu_M).
\label{eq:cond_y_M_M_odd}
\end{equation}
Similarly, $y_1$ is expressed in terms of $y_{M+1},y_{M+2},\dots$ as
\begin{equation}
y_1=X_1+\sum_{m=M+1}^\infty\frac{y_m}{m!}s_{\alpha(m-M,M-1)}(\mu_1,\dots,\mu_M).
\label{eq:y_1_expanded_M_odd}
\end{equation}
Now Eve's task is to minimize Eq. (\ref{eq:y_1_expanded_M_odd}) by adjusting $y_{M+1},y_{M+2},\cdots$
while maintaining inequality (\ref{eq:cond_y_M_M_odd}).
Note that both the relations are linear in $y_{M+1},y_{M+2},\cdots$, and thus the best configuration that
minimizes $y_1$ will be determined by their coefficients,
$s_{(m-M)}(\mu_1,\dots,\mu_M)/m!$ and $s_{\alpha(m-M,M-1)}(\mu_1,\dots,\mu_M)/m!$.
In fact, as we will show in Appendix \ref{app:Proof_Monotonically_Increasing},
the ratio of these two coefficients
\begin{equation}
K_m:=\frac{s_{\alpha(m-M,M-1)}(\mu_1,\dots,\mu_M)}{s_{(m-M)}(\mu_1,\dots,\mu_M)}
\label{eq:K_m_defined}
\end{equation}
increases monotonically with respect to $m$.
Hence the minimum value is achieved by maximizing as many $y_m$'s as possible with larger $m$'s
in such a way that is consistent with Eq. (\ref{eq:cond_y_M_M_odd}).
If the equality can be achieved in (\ref{eq:cond_y_M_M_odd}) for some configuration of $y_{M+1},y_{M+2},\dots$,
this amounts to finding $L(>M)$ such that $y_m=1$ for $m\ge L$,
$y_m=0$ for $M<m<L$, and $0\le y_L\le1$ for $m=L$, and also $y_M=0$ is satisfied.
On the contrary if the equality does not hold for any configuration,
$y_1$ is minimized when $y_n=1$ for all $n>M$.
Both these cases corresponds to $Z$ of Definition \ref{def:define_large_Z}.
Hence $Z$ thus obtained indeed gives the minimum of $y_1$ under our temporal constraints on $y_n$.

\begin{lemma}\label{lm:boundary_Z}
If $\mu_1,\dots,\mu_M\le1$,
then $0\le Z_n\le1$ is satisfied for all $n\le M$.
\end{lemma}

{\it Proof:}
Recall $0\le X_n\le 1$ when $\mu_i\le1$ from Lemma \ref{lm:boundary}.
Substituting $y_n=Z_n$ in (\ref{eq:difference_expanded}), we find for $n\le M$,
\begin{equation}
\frac1{n!}Z_n=\frac1{n!}X_n+(-1)^{M-n+1}\sum_{m=M+1}^\infty \frac{Z_m}{m!}s_{\alpha(m-M,M-n)}(\mu_1,\dots,\mu_M).
\label{eq:Z_n_expanded}
\end{equation}
Now since the Schur polynomial $s_{\alpha(m-M,M-n)}$ and $Z_m$ for $m>M$ being positive,
it is clear that $Z_n\le1$ for $M-n$ even,
and $Z_n\ge0$ for $M-n$ odd.

On the other hand, in order to show $Z_n\ge0$ for $M-n$ even
and $Z_n\le0$ for $M-n$ odd,
suppose we had $M-1$ constraints, say, of $\mu_1,\dots,\mu_{M-1}$ from the beginning,
and consider the corresponding $X$ and $w^{(m)}$,
which we will denote in what follows as $\bar{X}$ and $\bar{w}^{(m)}$.
Lemma \ref{lm:boundary} holds in this case as well and we have $0\le\bar{X}_n\le1$.
By definition, $\bar{X}$, as well as $Z$, are a solution to Eq. (\ref{eq:constraint_Q_positive})
for $i=1,\dots,M-1$.
Hence we can apply the same argument as in the previous paragraph,
using $\bar{w}^{(m)}$ and $\bar{X}$ this time,
and express $Z_n$ for $n\le M-1$ as
\begin{equation}
\frac1{n!}Z_n=\frac1{n!}\bar{X}_n+(-1)^{M-n}\sum_{m=M}^\infty \frac{Z_m}{m!}s_{\alpha(m-M-1,M-n-1)}(\mu_1,\dots,\mu_{M-1}).
\end{equation}
Again due to the positivity of the Schur polynomials and $Z_m$,
this shows $Z_n\ge0$ for $M-n$ even, and $Z_n\le0$ for $M-n$ odd.
This completes the proof.

\section{Conclusion}\label{sec:conclusion}
In this paper,
we presented an improved upper and lower bounds $X_n$, $Z_n$ for the asymptotic yield $y_n$
for the decoy state method using a finite number $M$ of decoy light intensities.
In particular if all the light intensities $\mu_i$ are less than or equal to one,
$X_n$, $Z_n$ are not only a lower or upper bound,
but in fact are the exact minimum or maximum.

Moreover, these $X_n$ and $Z_n$ can always be obtained by simple numerical calculation
by using Eq. (\ref{eq:y_1_explicit_form}), (\ref{eq:y_bar_expicit_form})
and by using the algorithm given in Sec. \ref{sec:algorithm_Z_n}.

\ 

\noindent{\large\bf Acknowledgment}

This work was supported by the project ``Research and Development on Quantum Cryptography
of the National Institute of Information and Communications Technology,
as part of Ministry of Internal Affairs and Communications of Japan's program
``R\&D on Quantum Communication Technology."



\appendix
\section{Properties of the Schur polynomials}\label{app:properties_Schur_poly}
The Schur polynomial $s_\lambda$ given in Definition \ref{def:Schur_polynomial}
can also be expressed as a sum of monomials as
\begin{equation}
s_\lambda(\mu_1,\dots,\mu_M)=\sum_T \mu_1^{t_1}\mu_2^{t_2}\cdots\mu_M^{t_M},
\label{eq:schur_poly_expanded}
\end{equation}
where $T$ denotes a semistandard Young tableaux on a Young diagram $\lambda$,
on which number $i\in\{1,\dots,M\}$ appears $t_i$ times (see, e.g., \cite{Fulton1, Fulton2}).
Semistandard tableaux are those having entries which are strictly increasing vertically
and weakly increasing horizontally \footnote{In some textbooks (e.g., \cite{Fulton2}),
a semistandard tableau is simply called a `tableau.'}.
For example, ${\scriptsize\Yvcentermath1\young(1123,2)}$ is semistandard whereas
${\scriptsize\Yvcentermath1\young(1234,1)}$ is not.
The monomial corresponding to the former tableau is $\mu_1^2\mu_2^2\mu_3$.
For $M=3$ and $\lambda =(2,1)= {\tiny\Yvcentermath1\yng(2,1)}$,
there are eight semistandard tableaux,
${\scriptsize\Yvcentermath1\young(11,2),\young(12,2),\dots,\young(23,3)}$,
and the Schur polynomial reads
\begin{eqnarray*}
s_{(2,1)}(\mu_1,\mu_2,\mu_3)&=&2\mu_1\mu_2\mu_3
+\mu_1\mu_2^2+\mu_2\mu_3^2+\mu_3\mu_1^2
+\mu_1^2\mu_2+\mu_2^2\mu_3+\mu_3^2\mu_1\\
&=&(\mu_1+\mu_2)(\mu_2+\mu_3)(\mu_3+\mu_1),
\end{eqnarray*}
which equals the one obtained from Definition \ref{def:Schur_polynomial}.

If $\mu_1,\dots,\mu_M>0$,
the polynomials $s_\lambda$ are always positive 
since the coefficient of each monomial is positive in Eq. (\ref{eq:schur_poly_expanded}).
In this case there is a simple upper bound 
\[s_\lambda(\mu_1,\dots,\mu_M)\le (\mu_{\rm max})^d\cdot s_\lambda(1,\dots,1)\]
with $d:=\sum_i \lambda_i$ and $\mu_{\rm max}=\max_i\mu_i$.
From this and using the formula
\[
s_{\lambda}(1,1,\dots,1)=\prod_{i<j}\frac{\lambda_i-\lambda_j+j-i}{j-i},
\]
(see, e.g., \cite{Fulton1,Fulton2}) we find
\begin{equation}
s_\lambda(\mu_1,\dots,\mu_M)\le (\mu_{\max})^d \prod_{i<j}\frac{\lambda_i-\lambda_j+j-i}{j-i}.
\label{eq:s_lambda_bound}
\end{equation}


\section{Proof of Lemma \ref{lm:w_completeness}}\label{app:proof_lemma_1}
For $m>M$, define $x^{(m)}=(x^{(m)}_1,x^{(m)}_2,\dots)$ as follows.
For $n=1,\dots,M$, let
\begin{equation}
x^{(m)}_n:=
(-1)^{M-n+1}n!\cdot\left|
\begin{array}{ccccccc}
\mu_1&\cdots&\mu_1^{n-1}&\mu_1^{n+1}&\cdots&\mu_1^M&\mu_1^m\\
\mu_2&\cdots&\mu_2^{n-1}&\mu_2^{n+1}&\cdots&\mu_2^M&\mu_2^m\\
\vdots&\vdots&\vdots&\vdots&\vdots&\vdots&\vdots\\
\mu_M&\cdots&\mu_M^{n-1}&\mu_M^{n+1}&\cdots&\mu_M^M&\mu_M^m
\end{array}
\right|
\label{eq:GVM1}
\end{equation}
and for $n=m$, let $x^{(m)}_m$ be
\begin{equation}
x^{(m)}_m:=m!\cdot
\left|
\begin{array}{ccc}
\mu_1&\cdots&\mu_1^M\\
\vdots&\ddots&\vdots\\
\mu_M&\cdots&\mu_M^M
\end{array}
\right|=m!D(\mu_1,\dots,\mu_M).
\label{eq:GVM2}
\end{equation}
All other elements of $x^{(m)}$ are zero.
Then it is easy to see that for $i=1,\dots,M$,
\[
\sum_{n=1}^\infty\frac{\mu_i^n}{n!} x^{(m)}_n=\sum_{n=1}^M\frac{\mu_i^n}{n!} x^{(m)}_n+\frac{\mu_i^m}{m!}x^{(m)}_m
=\left|
\begin{array}{cccc}
\mu_i&\cdots&\mu_i^M&\mu_i^m\\
\mu_1&\cdots&\mu_1^M&\mu_1^m\\
\mu_1&\cdots&\mu_2^M&\mu_2^m\\
\vdots&\vdots&\vdots&\vdots\\
\mu_M&\cdots&\mu_M^M&\mu_M^m
\end{array}
\right|=0.
\]
$w^{(m)}$ in Definition \ref{def:Delta_y} is expressed as $w^{(m)}_n=x^{(m)}_n/x^{(m)}_m$
and thus we have shown that $w^{(m)}$ is indeed the solution.

Next we prove that expansions in $w^{(m)}$ are possible.
For a given $\Delta y$, define $\Delta y':=\sum_{m=M+1}^\infty \Delta y_mw^{(m)}$
and consider $v:=\Delta y'-\Delta y$.
With $\Delta y$ and $\Delta y'$ both being a solution to Eq. (\ref{eq:def_Delta}),
$v$ is also a solution.
Then note that by definition $v_n=0$ for all $n>M$,
and thus $v_1,\dots v_M$ satisfy
\[
\left(
\begin{array}{ccc}
\mu_1&\cdots&\mu_1^M\\
\vdots&\ddots&\vdots\\
\mu_M&\cdots&\mu_M^M
\end{array}
\right)
\left(
\begin{array}{c}
v_1/1!\\
\vdots\\
v_M/M!
\end{array}
\right)=0.
\]
From this it follows $v_n=0$ for $n\le M$ as well,
due to the invertibility of the matrix on the left hand side.
Hence we have shown $\Delta y'=\Delta y$ and that any $\Delta y$ can be expanded with $w^{(m)}$.

In order to prove the uniqueness of the coefficients of $w^{(m)}$,
it suffices to show the linear independence of $w^{(m)}$.
This is obvious from the fact that for any $n>M$,
there is only one $w^{(m)}$ with a nonzero value in the $n$-th element,
i.e., $w^{(n)}$.


\section{$K_m$ is Monotonically Increasing in $m$}\label{app:Proof_Monotonically_Increasing}
\noindent{\it Proof:}
In this proof the variables of the Schur polynomials are always $\mu_1,\dots,\mu_M$,
and we will omit them for the sake of brevity.
It is immediate from Definition \ref{def:Schur_polynomial} that
$K_m$ can be rewritten as
\[
K_m=\left(\prod_{i=1}^M \mu_i\right)\frac{s_{(m-M-1)}}{s_{(m-M)}},
\]
and by using this we obtain
\begin{equation}
K_{m+1}-K_m=
\left(\prod_{i=1}^M \mu_i\right)
\frac{\left(s_{(m-M)}\right)^2-s_{(m-M+1)}s_{(m-M-1)}}{s_{(m-M)}s_{(m-M+1)}}.
\label{eq:K_m_difference}
\end{equation}
Multiplication of two Schur polynomials $s_\lambda$ and $s_\nu$ is especially simple
when the partition $\nu$ (or equivalently $\lambda$) consists of a single number
$\nu=(b)$.
That is,
\[
s_\lambda\cdot s_{(b)}=\sum_\rho s_\rho,
\]
where the sum is over all partitions $\rho$ that are obtained from $\lambda$ by adding $b$ boxes,
with no two in the same raw (see, e.g., Section 2.2 of Ref. \cite{Fulton2}).
Hence for $\lambda=(a)$ we have
\[
s_{(a)}\cdot s_{(b)}=
\sum_{c=0}^{\min(a,b)}s_{(a+b-c,c)},
\]
and from this it follows that the numerator of Eq. (\ref{eq:K_m_difference}) is positive;
$\left(s_{(m-M)}\right)^2-s_{(m-M+1)}s_{(m-M-1)}=s_{(m-M,m-M)}>0$.
Hence Eq. (\ref{eq:K_m_difference}) is also always positive,
meaning that $K_m$ is monotonically increasing.


\section{Upper bound on $L_0$}\label{seq:upper_bound_L_0}
For $\mu_i\le1$, $L_0$ can be bounded from above as follows.
If equality cannot hold in (\ref{eq:cond_y_M_M_odd}) for any configuration of $y$,
we have $L_0=M+1$
(cf. the argument below Eq. (\ref{eq:K_m_defined})).
On the contrary, if $Z_M=z_M(L_0,a_0)=0$ for some $(L_0,a_0)$, we have from Eq. (\ref{eq:z_M_expanded})
\begin{equation}
\frac{X_M}{M!}\le \sum_{m=L_0}^\infty\frac1{m!}s_{(m-M)}(\mu_1,\dots,\mu_M).
\end{equation}
Next using the upper bound of (\ref{eq:s_lambda_bound})
and applying a similar argument as in (\ref{eq:I_n_upperbound_origin}),
we find
\begin{eqnarray*}
\frac{X_M}{M!}&\le& \sum_{m=L_0}^\infty\frac1{m!}\frac{(m-1)!}{(m-M)!(M-1)!}\\
&=& \sum_{k=0}^\infty\frac1{(L_0+k)(L_0-M+k)!(M-1)!}\\
&=&\frac1{(M-1)!}\sum_{k=0}^\infty\frac1{k!}\frac{k!}{(L_0+k)(L_0-M+k)!}\\
&\le&\frac1{L_0(L_0-M)!(M-1)!}\sum_{k=0}^\infty\frac1{k!}\\
&=&\frac{e}{L_0(L_0-M)!(M-1)!}.
\end{eqnarray*}
As can be seen from (\ref{eq:y_n_extended}) $X_M$ is bounded from below as $X_M\ge q_M$,
and $L_0$ is upper bounded as
\begin{equation}
L_0(L_0-M)!\le\frac{Me}{X_M}\le\frac{Me}{q_M}.
\label{eq:L_0_bound}
\end{equation}

\end{document}